\documentclass[aps,prl,twocolumn,superscriptaddress,nofootinbib,
nobalancelastpage,nobibnotes]{revtex4}
\usepackage{epsfig}
\usepackage{graphicx}
\usepackage{caption}

\begin{document}

\title
{Quantum break in high intensity gravitational wave interactions }

\author{R. F. Sawyer}
\affiliation{Department of Physics, University of California at
Santa Barbara, Santa Barbara, California 93106}

\begin{abstract}
The lowest order amplitudes for [$ {\rm graviton+graviton \rightarrow photon +photon}]$ lead to cross-sections of order $G^2$, where $G$ is the gravitational constant. These are too small to be of any interest.
However in dense clouds of pure gravitons there are collective effects utilizing these same amplitudes that under the right circumstances can lead to copious production of photons.
\end{abstract}
\maketitle
 
 Confirming the quantum nature of the gravitational field puts us in the bind:

A. The cross-sections for individual gravitons on anything are too small to be observationally accessible.

B. Coherent many body systems of gravitons, as in a wave detectable by LIGO, are essentially classical fields, at least in current descriptions. And their detectable interactions are classical interactions.

Here we shall exhibit processes that escape these constraints.  They combine the advantages of explicit factors of $N_g$ from coherence, where $N_g$ is the number of gravitons in a mode, with an essential dependence on the quantization of gravity; and they depend linearly on amplitudes, not on cross-sections. 

The key here will be the fact that there exists a self-acting mechanism that can transform coherent classical clouds of gravitons into a similar construct of photons, and that produces a time scale proportional to  $G^{-1}$, rather than to $G^{-2}$ (as from cross-sections). Some elements are:

1. When we resolve the initial state into plane wave quanta that comprise the initial state (e.g. gravitons), the 3-momenta of the individual particles comprising the final state (photons and gravitons) are the same as in the initial state. 

2. The spatial and angular distributions within the initial graviton system must be such that, at least in some region, a tiny coherent mixing in amplitude of E\&M quanta with gravitons would grow exponentially in time, this in a conventional framework of ``mean-field" theory (or ``essentially classical" theory). 

3.  In the literature there are a number of examples of other systems at similar unstable equilibrium states in mean field theories (MF), with 
mechanisms whereby such instabilities are activated by quantum effects. They include:  Bose condensates of atoms in wells \cite{va}-\cite{va3} ;  polarization exchange processes in colliding photon beams  \cite{rfs1}, \cite{rfs4} ; axion decay into photons \cite{rfs8} ; cosmology \cite{cos2},\cite{cos3}, and the realization of a ``quantum speed limit" by a certain spin-lattice with infinite range ``x-y" model couplings \cite{qsl}. There is also a relation to ``fast neutrino flavor exchange" in the neutrino-sphere region in the supernova \cite{rfs2}-\cite{mir}, where the implicit $\hbar$ enabling the break is in the neutrino mass term. 

The two ``quantum break" examples cited above that involve continua of momentum states (the photon cloud and the neutrino cloud cases)
have a common qualitative feature that transcends their difference in statistics: they can be described as a transition in which momenta
of a swarm of particles stay exactly the same, but in which some discrete quality that we call polarization or flavor gets exchanged from beam to beam. 
For high initial occupancies, $N$, a typical scenario involves an extended gestation period followed by a very sudden transformation (in our case gravitons into photons).  This behavior  underlies the designation ``quantum break", and the basic calculation is of a quantum break time, $T_B$.
 \subsection{2. The basic amplitude}
   In all of the following, ``gr." will stand for ``graviton".
We need the non-vanishing invariant amplitudes for processes,  [${\rm gr.+\gamma  \rightarrow gr.+\gamma}$], as functions of the Mandelstam invariant variables
 $\mathcal{S},\mathcal{T},\mathcal{U}$ and the helicities \cite{sko}- \cite{grav2}. From eq.(49) in \cite{gris} we have,
\begin{eqnarray}
F_{2,1;2,1}=F_{-2,-1;-2,-1}=i {8 \pi G} ~{\mathcal{S}^2\over  \mathcal{T}}\, ,
\nonumber\\
\,
\nonumber\\
F_{2,-1;2,-1}=F_{-2,1;-2,1}=i {8 \pi G} ~{\mathcal{U}^2\over  \mathcal{T}}\, .
\label{gris1}
\end{eqnarray}
Crossing in the form of $t\leftrightarrow s$, $u\rightarrow{u}$ and the use of the helicity crossing relations from eq.(12) in \cite{gris}, gives us corresponding 
amplitudes for [gr.+gr.$\leftrightarrow \gamma+\gamma$],
\begin{eqnarray}
A_{-1,1;2,-2}=A_{1,-1;2,-2}=i {8\pi G } ~  {\mathcal{T}^2 \over  \mathcal{S}}\,,
\nonumber\\
\,
\nonumber\\
A_{-1,-1;2,2}=A_{1,1;-2,-2}=i {8\pi G} ~ {\mathcal{U}^2 \over  \mathcal{S}}\,.
\label{amps}
\end{eqnarray}
In (\ref{gris1}) for the scattering reaction, $\mathcal{T}$ is the square of the 4-momentum transfer, initial to final graviton. 
We shall work out our first example here taking the initial state as two clashing beams of gravitons each with helicity wave function, 
$[ | 2\rangle +| -2\rangle ]2^{-1/2}$. 
Using the amplitudes ({\ref{amps}), we see that this leads to the two photon state 
\begin{eqnarray}
\mathcal{S}^{-1}\Bigr\{  (\mathcal{T}^2+\mathcal{U}^2)  \Bigr [|1\rangle +|-1\rangle \Bigr ] \Bigr [|1\rangle +|-1\rangle \Bigr ]
\nonumber\\
+(\mathcal{T}^2- \mathcal{U}^2) \Bigr [ |1,-1 \rangle +|-1,1 \rangle  -|1,1 \rangle -|-1,-1 \rangle \Bigr ]\Bigr \}
\nonumber\\
\,
\label{amp2}
 \end{eqnarray}

We note that the second line in (\ref{amp2}) will vanish under Bose symmetrization.
Thus with this choice of initial states we avoid a multi-channel calculation in helicity space. (In sec.5 we discuss multichannel calculations in a general way.)
Next we note that
the dynamics of the long term coherence necessary for the break into $\gamma$'s requires  either ${\bf p,q \rightarrow  p,q } $ or ${\bf p,q \rightarrow\, q,p}$ in the basic two-particle reaction. This requires either ($\mathcal{T}=0$, $\mathcal{U}=-\mathcal{S}$), or ($\mathcal{T}=-\mathcal{S}$, $\mathcal{U}=0$).  From (\ref{amp2}) we obtain the value $i 8\pi G \mathcal{S} $ for the amplitude in either case.
We now turn this into an effective Hamiltonian for computation of time development of multi-particle systems. 
The lab system that we choose for this purpose should be dictated by the distribution of masses in the system that generates the initial gravitational waves. Here we think of an extended system, such as a black hole binary roughly at rest in a particular lab system.
Through its own instabilities it produces gravitational waves, which in our picture interact with each other to produce photons.
The system itself will be contained in a periodic box of size much greater than the particle wavelengths, volume $V$, and also little larger than the light travel distance over the period that we need to follow it to see transitions. Of course the number of particles  contained will be enormous. The effective Hamiltonian 
for the graviton pair transition to a photon pair, with momenta p and q, then is given by,

\begin {eqnarray}
H_{\rm g,\gamma}= {  2\pi G \mathcal{S}\over {\bf |p| \, |q| }V} \Bigr [ a^\dagger b ^\dagger c \, d+ a \,b c^\dagger\, d^\dagger \Bigr ]
\nonumber\\
= { 2\pi G \mathcal{S} \over  {\bf |p| \, |q|} V} \Bigr [ \sigma_+ \tau_+ +\sigma_- \tau_- \Bigr ]\,,
\nonumber\\
\label{ham1}
 \end{eqnarray}
in terms of the operators that create and annihilate single particle momentum states,  $a$ and $b$ for the single gravitons
with respective momenta ${\bf p}$ and ${\bf q}$; $c$ and $d$ similarly for the photons; or, in the second form, in terms of the operators
$\sigma_+=a^\dagger c$, which changes species for the ${\bf p}$ beam , and $\tau_+=b^\dagger d $ that refers to ${\bf q}$. Note that the frame dependence of the other factors is entirely
in $(\,{\bf |p| \, |q| }V \,)^{-1} \mathcal{S}=2 (1-\cos \theta_{p,q})V^{-1}$, peaked in favor of opposing momenta and vanishing for parallel momenta; and that 
$ \theta_{p,q}$
is not a scattering angle, but rather an angle of incidence in the lab system.

Our most primitive picture will be one in which somewhere in the interior of the turmoil two high occupancy beams, each with high occupancy $N=n V$, meet head-on, $\cos \theta_{p,q}=-1$. Multiple beam solutions allowing energy and angular distributions are discussed in sec. 5.
The wave function for the state with $N$ gravitons in each of the ${\bf p}$ and ${\bf q}$ beams now lies in an $N+1$ dimensional space with  
with basis vectors that describe states with $N-k+1$ gravitons and $k-1$ photons in each of the two beams, where $k=1,2...N+1$. The non-vanishing matrix elements that enter are,
\begin{eqnarray}
\langle k-1 |\sigma_+ \tau_+ |k\rangle=k (N-k+1)\,,
\nonumber\\
\langle k |\sigma_- \tau_-|k-1\rangle=k (N-k+1)\,.
\label{equal}
\end{eqnarray}

We solve for the wave function $|\Psi(t)\rangle$ in the $N+1$ dimensional subspace and calculate
\begin{eqnarray}
 \zeta(t)=N^{-1}\langle \Psi(t)| (1/2+\sigma_3/2)|  \Psi(t)\rangle  \,.
 \end{eqnarray}
In fig.1 the dashed curves plot the evolution of $\zeta(t)$, the probability of an initial graviton to remain a graviton, as a function of scaled time $s$ for values $N=64, 256, 1024$, where,
\begin{eqnarray}
  s=8 \pi G n\,t \,,
  \label{rescale}
 \end{eqnarray}
and $n=N/V$ the $t=0$ number density of gravitons in either beam.

The $N$'s in this calculation are pathetically small; but the plotted results are of prime importance in checking the mean-field methods to come. 
\begin{figure}[h] 
 \centering
\includegraphics[width=2.5 in]{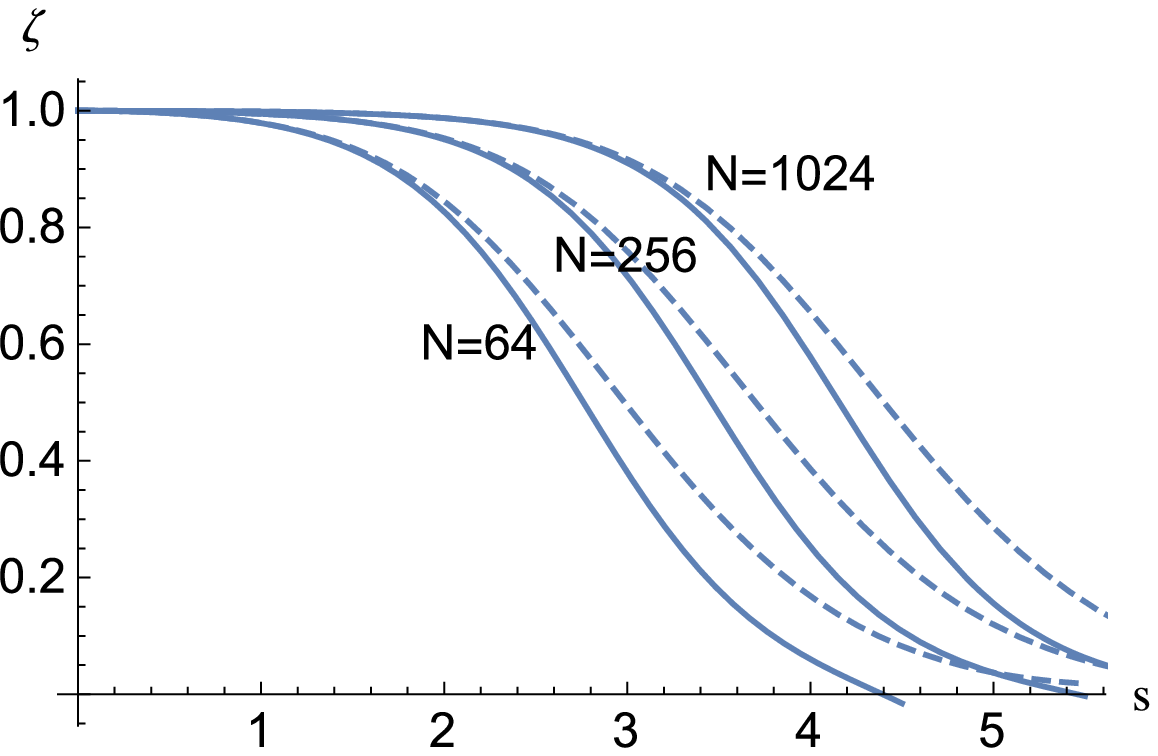}
 \caption{ \small }
Dashed curves: retention probability $\zeta (s) $in the wave function solution. Solid curves: the same in the corresponding MMF solutions, with their  $N$ values identifiable by their coalescence with the corresponding dashed curves for small scaled time $s$. $\zeta=1$ indicates 100\% gravitons.
\label{fig. 1}
\end{figure}
\subsection{3. MMF approximation}
We return to the bilinears defined earlier, $\vec \sigma$ for the ${\bf p}$ stream and  $\vec \tau$ for the ${\bf q}$ stream, 
and define $X=\sigma_+ \tau_+$, $Y=\sigma_+ \sigma_- + \tau_+ \tau_-$ . We rename  $\sigma_3= \tau_3=Z$ (their equality being chosen in an initial condition, and then maintained throughout). The operators $\vec \sigma$, and $\vec \tau$ have the commutation relations of Pauli matrices, but actually are angular momentum matrices  (times 2) of dimension $N$;  we use only their commutators in the following.
The Hamiltonian is now
\begin{eqnarray}
H_{\rm g,\gamma}={8 \pi G \over V}[X+X^\dagger] \,.
\end{eqnarray}
Using commutators to obtain the Heisenberg equations of motion
 we obtain, 

\begin{eqnarray}
i \dot X ={8 \pi G \over V} (Z Y-Z^2) \, ,
\nonumber\\
i \dot Y={16 \pi G \over V}  Z ( X^\dagger-X) \, ,
\nonumber\\
i \dot Z ={16 \pi G \over V} (X-X^\dagger )\,.
\label{eom3}
\end{eqnarray}
The $Z^2$ term in the first equation comes from a second commutation to get operators into the correct order; implicitly it carries an additional power of $\hbar$ and is the source of the ``quantum break" to come.
Our modified mean field method (MMF) is to replace each of the operators $X, Y, Z$ in (\ref{eom3}) by its expectation value in the medium, thus implicitly assuming that, e.g, $\langle Z Y\rangle= \langle Z \rangle \langle Y\rangle$. 

Next we do a rescaling in which each one of the single particle operators $a, b, c, d, a^\dagger...$ is redefined by extracting a factor of $N^{1/2}$, so that $x=X/N^2$, $y=Y/N^2$, $z=Z N$ and at the same time defining $n=N/V$, the number density, and the scaled time variable, $s$, according to (\ref{rescale}).
 
 The rescaled equations are,
\begin{eqnarray}
i {d x \over ds} =zy-z^2/N \,,
\nonumber\\
i {d y \over ds} =2 z ( x^\dagger-x) \,,
\nonumber\\
i {d z \over ds} =2 (x-x^\dagger ) \,.
\label{mmf}
\end{eqnarray}
Solutions are shown as solid lines in fig. 1.  The zero-parameter fit of the complete solutions to the MMF solutions
over the time required for  20\% of the gravitons to transform to photons gives us some degree of confidence in our later use of MMF for astronomically greater numbers of particles, at least at early times.  In fig.2 we show solutions to (\ref{mmf}) for geometrically spaced higher values of $N$.   The equal spacings indicate the behavior of the turnover time $T_B\sim (8 \pi G n)^{-1} \log N$.  The comparative suddenness of the transition
earns it the designation ``quantum break". 
 \begin{figure}[h] 
 \centering
\includegraphics[width=2.5 in]{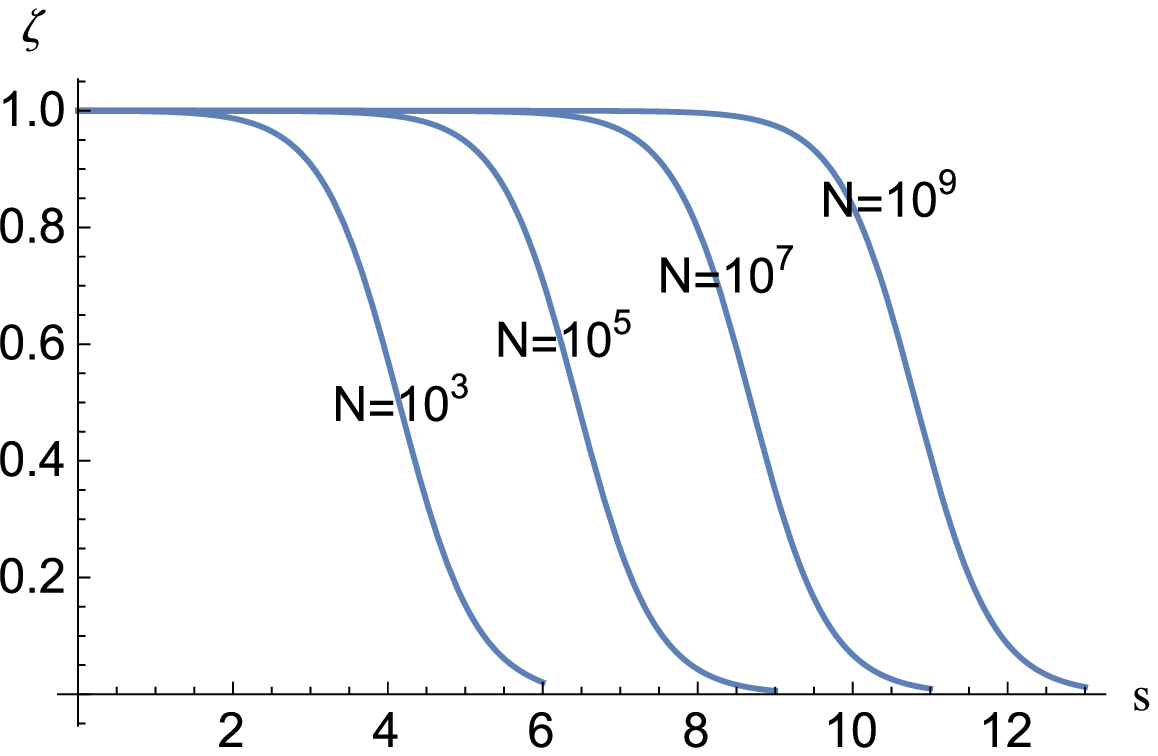}
 \caption{ \small }
The same as fig.1, except only the MMF solutions, for a series of higher values of N.
\label{fig. 2}
\end{figure}

 \subsection{4. Instability condition.}
In mean field language, the  turn-over in time proportional to $G^{-1}$, rather than $G^{-2}$ as would have been expected from cross-sections, is an instability of the classical equilibrium state with clashing gravitational waves. In the quasi-stable initial state with $z=1,x=0, y=0$, the term $z^2/N$ in the $x$ equation, with it's implicit $\hbar$ factor, is what drives the instability. Dropping this term, taking $z=1$ elsewhere in the equations and looking at now linearized equations for $\Delta[x(s)],\Delta[x^\dagger (s)] ,  \Delta[y(s)]$ we find a 3$\times$3 response matrix with eigenvalues  (in units of inverse scaled time) just given by $0, \pm 2$. This gives us exponential behavior that fits the plotted plunges shown in fig. 2., although, by itself, no hint of the timing of the plunge.

In trading a graviton for a photon we implicitly assumed
that when the trade was finished the system energy was unchanged. But if the photon's energy shift from its interactions in the medium is different from the graviton's (speaking loosely) it would introduce an incoherence that could doom the instability. In our formalism we now need to study the effects of the ${\rm gr. +\gamma \rightarrow gr. +\gamma} $ on the evolution of our variables $X,Y,Z$. Here the singular factor $\mathcal{T}^{-1}$ in the scattering amplitudes (\ref{gris1}) is a warning, screaming ``long range" when ``local" is our operational basis for everything. We also require the graviton-graviton scattering amplitude, given in eq.(40) in \cite{gris}. The photon-photon amplitude does not enter as long as we are concentrating on the early time instability, when there are next to no photons. We also need to be aware that the parameter $c$ in the amplitudes given in \cite{gris} changes its definition from $i/4$ to $-i/4$ in going from the [$\gamma+{\rm gr.} $] case to the [$\rm{gr.}+{\rm gr.} $] case.

We have studied these corrections in depth only for the case of our simplest and most potent configuration of two beams colliding head-on. 
The new terms in the Hamiltonian in our notation are now proportional to $(a^\dagger a d^\dagger d+b^\dagger b c^\dagger c)$, for the 
[ $\gamma+{\rm gr.}] $ part and to $(a^\dagger a b^\dagger b+c^\dagger c d^\dagger d)$ for the [$\rm{gr.}+{\rm gr.} $] part. The resulting changes to the evolution equations (\ref{eom3}) miraculously cancel between the pieces generated by these two new terms. Thus we do not need to revisit the instability condition after all, at least for our simplest configuration.
 
We remark that an ordinary mean-field theory based on the field variables $a,b,c,d$, rather than on the quartics that we have employed, also leads to growing modes in a linearized instability analysis. But in this approach nothing happens (to order $G$) when we start with a pure graviton state. The system is in unstable equilibrium at this classical level.
\subsection{5. Multiple beams}
We replace our two beams ($\vec \sigma $ and $\vec \tau$ ) by $N_b$ beams $\vec \sigma_j$, for $j=1..N_b$, at different angles and with the  effective interaction
\begin{eqnarray}
H_{\rm eff}={ 4 \pi G \over  V}\sum_{j,k}^{N_b} [\sigma^{j}_+ \sigma_+^{k} +\sigma_-^{j} \sigma_-^{k}] \lambda_{j,k} \,,
\nonumber\\
\label{hamx}
\end{eqnarray}
where $\lambda_{j,k} =(1-\cos \theta_{j} \cos \theta_{k})$
and $\cos \theta_j$'s are uniformly distributed in the interval \{-1,1\}, the best to simulate an isotropy in the whole distribution, if each of the $N_b$ rays has $N/N_b$ occupancy.
 We define the quartic variables, 
 \begin{eqnarray}
X_{l,m}=\sigma_+^l \sigma_+^m ~~,~~Y_{l,m}=\sigma^l_+ \sigma_-^m \,.
 \nonumber\\
 \end{eqnarray} 
 The Heisenberg equations of motion for the system are, after rescaling,
\begin{eqnarray}
 &i {d \over ds} X_{r,m}=g_1\Bigr[ Z_m \sum _k^{N_b}\lambda_{r,k} Y_{m,k} +Z_r \sum_k^{N_a}\lambda_{m,k}  Y_{k,m}-
 \nonumber\\
& \lambda_{r,m} N^{-1} Z_m Z_r \Bigr ] \,,
\label{final1}
 \end{eqnarray}
 
  \begin{eqnarray}
  i {d \over ds}  Y_{r,m}=2 g_1 \Bigr [-Z_m \sum _k^{N_b} \lambda_{k,m}X_{r,k} +Z_r \sum_k^{N_b}\lambda_{k,r} X^*_{m,k}\Bigr ] \, ,
  \nonumber\\
  \,
  \label{final2}
 \end{eqnarray}
  
 \begin{eqnarray}
i{d \over ds}  Z_r=2 g_1 \sum_k^{N_b} \lambda_{r,k} (X_{r,k}-X_{r,k}^* )\, .
\label{final3}
 \end{eqnarray}

The time has been scaled here in much the same way as in our original two-beam $(\sigma, \tau)$ model. The total number density $n$ is the same in the two calculations. In the rescaling of the operators $ X, Y, Z$, however, we have used factors of $N/N_b$ where we used $N$ previously.
In consequence, the scaled coupling constant became  $g_1= N_b/N$ in (\ref{final3}) instead of unity as before. In fig. 3 we plot the results for the usual (graviton) retention variable $\zeta$ now for fixed $N =10^7$ as a function of the number of beams in the division, taking the three values $N_b=3,9,27$, (the last of which is  the highest that Mathematica will take us).

 \begin{figure}[h] 
 \centering
\includegraphics[width=2.5 in]{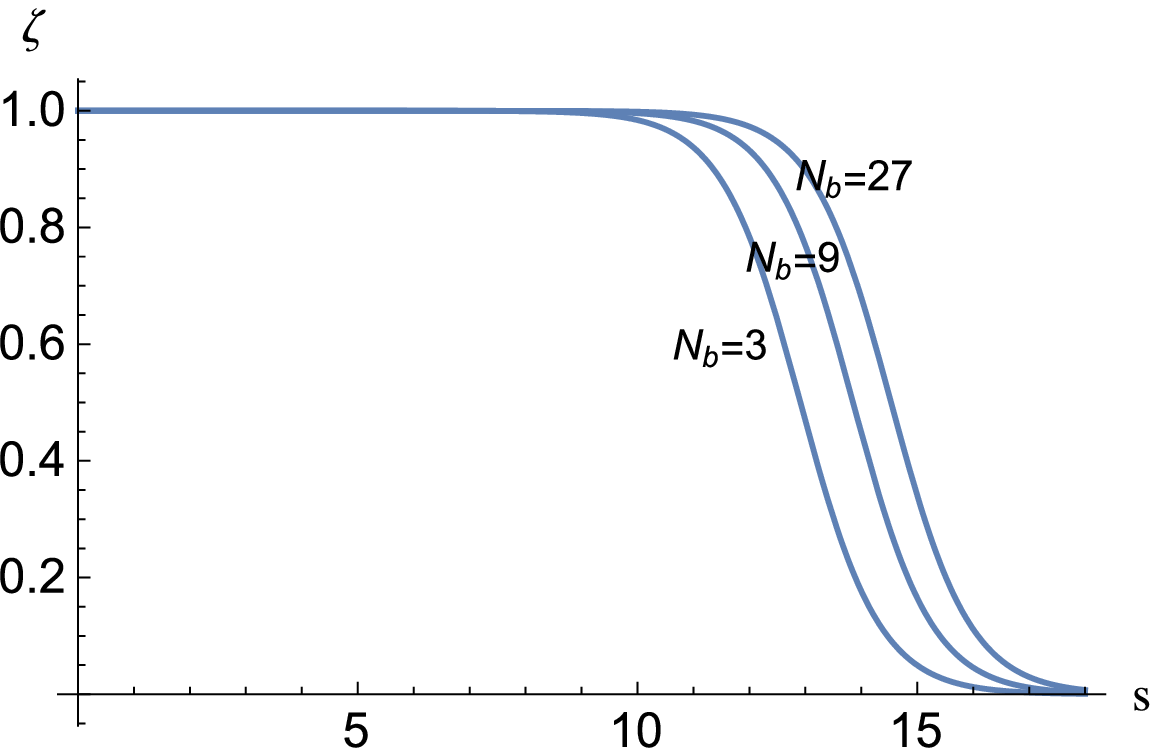}
 \caption{ \small }
The same as fig.2, but based on (\ref{final3}) with $N_b=3,9,27$, $N=10^7$
\label{fig. 3}
\end{figure}

With a finer and finer subdivision (larger $N_b$) it appears that that we would reach a limiting value for the time of the dive, and that it that is in the neighborhood of $1.5 \times$ the time shown in the $N=10^7 $ plot of fig. 2 for the ``two beams head-on" case.

Another use we can make of (\ref{final1})-(\ref{final3}) is to take the initial state to consist of half of the initial gravitons moving in the $+{\bf \hat z}$
direction and the other half moving in the $-{\bf \hat z}$ direction, so that $\lambda_{j,k}=2$ or zero for all $j,k$. Suppose that the
quanta in the sub-beams are distinguished by momentum magnitudes $p_j$ (or $q_j$).  Looking then at the solutions to (\ref{final3}) where the total initial graviton number $N$ is subdivided into groups with occupancy $N/ N_b$,  we might expect that for large values of $N_b$ we would lose big $\sqrt N$ factors that are characteristic of the matrix elements of a single annihilation operator in a classical coherent state. Note that in (\ref{final3}) $N$ enters explicitly in the quantum term in the $\dot X$ equation, and implicitly in the factor $g_1$, whereas $N_b$ enters in $g_1$ and in the sums. We compute examples with fixed $N=3\times10^6$ and $N_b$ ranging from 
2 to 30. Their break curves are absolutely identical over that range.  Thus we can have the same coherent phenomena in flows that are completely fragmented in absolute momentum; with whatever phase relations obtain among the components. And our claim in the introduction that neutrino clouds can show similar behavior (with the translation $[\gamma, {\rm gr.}]\rightarrow$``flavor") should seem less bizarre.

On a different aspect of the multi-beam case we note that exact energy conservation is an over-constraint. Our method was to start at $t=0$ with momentum states obeying periodic boundary conditions on a box of volume $V$, and we have chosen an initial state that would indeed have been an energy state in an absence of interaction. We calculate only up to the break time $T_B$ and in 3D there is a whole near-continuum of other final states available that can cohere in the collective effect, having unperturbed energies that differ from that of the initial state by $ \Delta E< T_B^{-1}$. These will only reduce the argument of the logarithm in the above estimate of $T_B$, leaving the coefficient the same, as was shown in the parallel axion calculation of \cite{refx}

\subsection{6. Discussion}

It may be surprising that, in principle, collisions of high intensity gravitational waves can produce photons on a time scale many orders of magnitude less than that estimated from graviton-graviton cross-section times number density. It is somewhat like an index of refraction process in which, say, a direction of polarization of a given momentum photon changes, in response to some non-isotropy of the medium, but it is much more complicated in that our gravitons must change two at a time, and that back-reactions are an essential ingredient as well.  An unusual feature
of the mechanics is how much it likes low frequencies, as first became apparent in the effective Hamiltonian (\ref{ham1}), which is
completely independent of frequency. Thus for a given energy density we would get the most action when the wave lengths are longest.

Since the black hole merger process is the one thought to be responsible for at least some of the LIGO events, our first estimate of the real-world possibilities for our effects will take it's numbers from there . The idea is that the produced gravitons in a region substantially larger than the horizon size would have time to interact for a time of $10^{-2}$ seconds or so, while the mechanisms of this paper produced photons. During this time we take the average power production in gravitational waves within the region to be $3\times 10^{56} $ ergs/sec. and assumed a wavelength centered at $10^7$ cm. 
Our estimate for a transformation time, given the implied graviton density in the region is then about $10^{-1 }\log_{10} N$ sec.  We miss relevance by a factor of ten even if the logarithm is ignored, whereas in the above scenario the direct estimate of the logarithm is about a factor of 30, but the factor will be diminished by the effects noted at the end of section 5. 

In any case we do not know if the black hole merger will turn out to be the best venue for our effects. There are many other species of ultra-energetic events out there, it now appears. Probably there are better methods than those we have used here for exploring these problems. If not, much could be done by a group with better access to computational power. In addition, as in any application of our results, there would be a tension between being close enough to strong fields for the production of the high densities of gravitational waves, and far enough away to do estimates based on keeping only the interactions treated in this paper. Subsequent to the completion of this work our attention was drawn to the work reported in refs. \cite{js1} and \cite{js2}. The approaches are totally different, and the domains of applicability appear to us to be different, but the central questions are closely related

It is a pleasure to thank Mark Srednicki  and Arkady Vainshtein for illuminating conversations.

\end{document}